\PassOptionsToPackage{unicode}{hyperref}
\PassOptionsToPackage{hyphens}{url}
\PassOptionsToPackage{dvipsnames,svgnames,x11names}{xcolor}
\documentclass[12pt]{article}

\usepackage{amsmath,amssymb}
\usepackage{iftex}
\ifPDFTeX
  \usepackage[T1]{fontenc}
  \usepackage[utf8]{inputenc}
  \usepackage{textcomp} 
\else 
  \usepackage{unicode-math}
  \defaultfontfeatures{Scale=MatchLowercase}
  \defaultfontfeatures[\rmfamily]{Ligatures=TeX,Scale=1}
\fi
\usepackage{lmodern}
\ifPDFTeX\else  
\fi
\IfFileExists{upquote.sty}{\usepackage{upquote}}{}
\IfFileExists{microtype.sty}{
  \usepackage[]{microtype}
  \UseMicrotypeSet[protrusion]{basicmath} 
}{}
\makeatletter
\@ifundefined{KOMAClassName}{
  \IfFileExists{parskip.sty}{%
    \usepackage{parskip}
  }{
    \setlength{\parindent}{0pt}
    \setlength{\parskip}{6pt plus 2pt minus 1pt}}
}{
  \KOMAoptions{parskip=half}}
\makeatother
\usepackage{xcolor}
\makeatletter
\ifx\paragraph\undefined\else
  \let\oldparagraph\paragraph
  \renewcommand{\paragraph}{
    \@ifstar
      \xxxParagraphStar
      \xxxParagraphNoStar
  }
  \newcommand{\xxxParagraphStar}[1]{\oldparagraph*{#1}\mbox{}}
  \newcommand{\xxxParagraphNoStar}[1]{\oldparagraph{#1}\mbox{}}
\fi
\ifx\subparagraph\undefined\else
  \let\oldsubparagraph\subparagraph
  \renewcommand{\subparagraph}{
    \@ifstar
      \xxxSubParagraphStar
      \xxxSubParagraphNoStar
  }
  \newcommand{\xxxSubParagraphStar}[1]{\oldsubparagraph*{#1}\mbox{}}
  \newcommand{\xxxSubParagraphNoStar}[1]{\oldsubparagraph{#1}\mbox{}}
\fi
\makeatother

\providecommand{\tightlist}{%
  \setlength{\itemsep}{0pt}\setlength{\parskip}{0pt}}\usepackage{longtable,booktabs,array}
\usepackage{calc} 
\usepackage{etoolbox}
\makeatletter
\patchcmd\longtable{\par}{\if@noskipsec\mbox{}\fi\par}{}{}
\makeatother
\IfFileExists{footnotehyper.sty}{\usepackage{footnotehyper}}{\usepackage{footnote}}
\makesavenoteenv{longtable}
\usepackage{graphicx}
\makeatletter
\def\maxwidth{\ifdim\Gin@nat@width>\linewidth\linewidth\else\Gin@nat@width\fi}
\def\maxheight{\ifdim\Gin@nat@height>\textheight\textheight\else\Gin@nat@height\fi}
\makeatother
\setkeys{Gin}{width=\maxwidth,height=\maxheight,keepaspectratio}
\makeatletter
\def\fps@figure{htbp}
\makeatother

\makeatletter
\@ifpackageloaded{caption}{}{\usepackage{caption}}
\AtBeginDocument{%
\ifdefined\contentsname
  \renewcommand*\contentsname{Table of contents}
\else
  \newcommand\contentsname{Table of contents}
\fi
\ifdefined\listfigurename
  \renewcommand*\listfigurename{List of Figures}
\else
  \newcommand\listfigurename{List of Figures}
\fi
\ifdefined\listtablename
  \renewcommand*\listtablename{List of Tables}
\else
  \newcommand\listtablename{List of Tables}
\fi
\ifdefined\figurename
  \renewcommand*\figurename{Figure}
\else
  \newcommand\figurename{Figure}
\fi
\ifdefined\tablename
  \renewcommand*\tablename{Table}
\else
  \newcommand\tablename{Table}
\fi
}
\@ifpackageloaded{float}{}{\usepackage{float}}
\floatstyle{ruled}
\@ifundefined{c@chapter}{\newfloat{codelisting}{h}{lop}}{\newfloat{codelisting}{h}{lop}[chapter]}
\floatname{codelisting}{Listing}

\makeatother
\makeatletter
\@ifpackageloaded{caption}{}{\usepackage{caption}}
\@ifpackageloaded{subcaption}{}{\usepackage{subcaption}}
\makeatother

\ifLuaTeX
  \usepackage{selnolig}  
\fi
\usepackage[]{natbib}
\usepackage{bookmark}

\IfFileExists{xurl.sty}{\usepackage{xurl}}{} 
\usepackage{amsthm,bm,physics,multirow,diagbox}
\usepackage{algorithm}
\usepackage{algorithmicx}
\usepackage{algpseudocode}

\newtheorem{theorem}{Theorem}
\newtheorem{lemma}{Lemma}
\newtheorem{proposition}{Proposition}

\theoremstyle{remark}
\newtheorem{remark}{Remark}

\theoremstyle{definition}

\newtheorem{assumption}{Assumption}

\newcommand{\argmin}{\operatornamewithlimits{arg\,min}}

\renewcommand{\Pr}{\mathbb{P}}				
\newcommand{\E}{\mathbb{E}}					

\newcommand{\spt}[1]{\operatorname{supp}(#1)}	
\newcommand{\inprod}[2]{\langle #1, #2 \rangle}	

\usepackage{xr-hyper}
\usepackage{hyperref}
\makeatletter
\newcommand*{\addFileDependency}[1]{
\typeout{(#1)}
\@addtofilelist{#1}
\IfFileExists{#1}{}{\typeout{No file #1.}}
}
\makeatother

\hypersetup{
  bookmarksnumbered=true,
  colorlinks=true,
  linkcolor=NavyBlue,		     
  citecolor=ForestGreen,       
  urlcolor=magenta	     	 
}

\begin{document}

\def\spacingset#1{\renewcommand{\baselinestretch}%
{#1}\small\normalsize} \spacingset{1}


\newcommand{\anon}{1}

\if1\anon
{
  \title{\bf Multi-transport Distributional Regression}
  \author{Yuanying Chen\\
  School of Statistics and Data Science,\\
    Shanghai University of Finance and Economics\\
    Tongyu Li\\
    Department of Statistics and Data Science,\\
    National University of Singapore\\
    Yang Bai\\
    School of Statistics and Data Science,\\
    Shanghai University of Finance and Economics\\
    Zhenhua Lin\thanks{Corresponding author: \texttt{linz@nus.edu.sg}.}\\
    Department of Statistics and Data Science,\\
    National University of Singapore}
  \date{}
  \maketitle
} \fi

\if0\anon
{
  \bigskip
  \bigskip
  \bigskip
  \begin{center}
    {\LARGE\bf Title}
\end{center}
  \medskip
} \fi

\bigskip
\begin{abstract}
We study distribution-on-distribution regression problems in which a response distribution depends on multiple distributional predictors. Such settings arise naturally in applications where the outcome distribution is driven by several heterogeneous distributional sources, yet remain challenging due to the nonlinear geometry of the Wasserstein space.
We propose an intrinsic regression framework that aggregates predictor-specific transported distributions through a weighted Fréchet mean in the Wasserstein space. The resulting model admits multiple distributional predictors, assigns interpretable weights quantifying their relative contributions, and defines a flexible regression operator that is invariant to auxiliary construction choices, such as the selection of a reference distribution.
From a theoretical perspective, we establish identifiability of the induced regression operator and derive asymptotic guarantees for its estimation under a predictive Wasserstein semi-norm, which directly characterizes convergence of the composite prediction map. 
Extensive simulation studies and a real data application demonstrate the improved predictive performance and interpretability of the proposed approach compared with existing Wasserstein regression methods.
\end{abstract}

\noindent%
\vfill

\newpage
\spacingset{1.8} 

\section{Introduction}
Distributional data analysis has attracted increasing attention as a statistical framework for analyzing data represented in the form of probability distributions, such as histograms, densities, and quantile functions. 
Rather than reducing such objects to scalar or vector summaries, this line of work focuses on statistical inference and modeling at the distributional level, allowing methods to capture richer structural information. 
Recent overviews of this field, such as \citet{PETERSEN2022model}, highlight its broad applicability across a wide range of domains, including demography \citep{hron2016simplicial,Bigot2017GPCA}, environmental science \citep{nerini2007classifying}, biomedical studies \citep{2016Functional}, and financial econometrics \citep{kokoszka2019forecasting}.

Within this framework, regression analysis plays a central role. 
In many contemporary applications, both predictors and responses are probability distributions, leading to distribution-on-distribution regression problems \citep{ghodrati2022distribution,Chen2023Wasserstein,zhu2023geodesic,ghosal2025distributional}. 
Such settings arise in a variety of domains, including mortality forecasting, where future age-at-death distributions depend on multiple historical population distributions; climate science, where regional outcome distributions are driven by several distributional climate forcings; and physiological signal analysis, where responses aggregate information from heterogeneous distributional sources. 
A defining feature of these problems is that the response distribution is influenced simultaneously by multiple predictor distributions, making multi-distributional regression practically important and calling for regression frameworks that can accommodate multiple distributional inputs.

A central difficulty in distribution-on-distribution regression stems from the absence of a linear structure in the Wasserstein space. 
This non-Euclidean structure renders classical linear regression tools inapplicable and has motivated a range of alternative methodological approaches. 
One line of work embeds distributions into Hilbert spaces through suitable transformations \citep{2016Functional, chen2019lqd}, while another exploits tangent-space linearizations of the Wasserstein manifold to perform regression in locally linearized spaces \citep{Chen2023Wasserstein, Zhang2022}. 
Although these approaches can be effective, they may distort the intrinsic geometry of the Wasserstein space or rely on local constructions that complicate interpretation and extension. 
More recently, optimal transport (OT)--based regression models \citep{ghodrati2022distribution, zhu2023autoregressive, ghodrati2024distributional} have been proposed that directly relate predictor and response distributions through transport maps, yielding intrinsically defined regression operators that preserve the underlying Wasserstein geometry and admit transparent interpretation in the distributional domain.

Despite these advances, extending intrinsic Wasserstein regression models to settings with multiple predictor distributions remains challenging. 
The nonlinear geometry of the Wasserstein space complicates the definition of aggregation operations that are simultaneously well-defined, interpretable, and commutative, and naive extensions may suffer from structural restrictions. 
Recent work has begun to address these issues. 
For instance, geodesic optimal transport (GOT) regression \citep{zhu2023geodesic} defines addition and scalar multiplication along Wasserstein geodesics, but imposes restrictive structural constraints and yields estimates that depend on predictor ordering. 
Alternative constructions based on parallel transport \citep{chen2024distribution} alleviate some of these limitations, yet challenges remain in achieving a flexible and interpretable framework for multivariate distribution-on-distribution regression.

To address these gaps, we propose a new intrinsic distribution-on-distribution regression framework based on weighted Fr\'echet means in the Wasserstein space. 
At a structural level, the proposed model represents the conditional regression operator as a weighted Fréchet mean of predictor-specific transported distributions. Each predictor enters the model through a pair consisting of a transport map and an associated scalar weight, which together characterize how the predictor distribution deforms and contributes to the response distribution. A fixed reference distribution is introduced to serve as an anchor, playing a role analogous to an intercept in classical regression. 
Importantly, although different choices of the reference distribution lead to different individual transport maps, the resulting regression operator and predictions remain unchanged, ensuring that the model is free of auxiliary construction dependence. This formulation yields a flexible and extensible regression operator that naturally accommodates multiple predictor distributions, while preserving interpretability and avoiding dependence on predictor ordering.

Our approach offers three key advantages: 
\begin{enumerate}
\tightlist
\item \emph{Flexibility.}
The model provides a flexible characterization of predictor-dependent transformations. In particular, the transformation maps are not restricted to geodesics toward Fréchet means, enabling a richer class of distributional shape deformations.
\item \emph{Extensibility.}
The introduction of a reference distribution facilitates a natural extension from single to multiple predictors while preserving identifiability. The reference distribution may be chosen freely, provided that it is fixed and known.
\item \emph{Interpretability.}
The regression operator admits a decomposition into transport maps and scalar coefficients constrained to lie in the unit simplex, providing a clear and interpretable quantification of predictor effects.
\end{enumerate}

From a theoretical perspective, we establish identifiability of the induced regression operator and derive asymptotic guarantees for the proposed M-estimator under a predictive Wasserstein semi-norm. Under this semi-norm, the population risk admits a global quadratic expansion around the true regression operator, providing a strong identification structure despite the infinite-dimensional and non-Euclidean nature of the parameter space. This framework yields consistency and convergence rates that are directly aligned with predictive accuracy, without requiring componentwise identifiability of individual transport maps or weights. 
The resulting convergence behavior exhibits a two-stage structure. When the regression operator is estimated from idealized distributions, the estimator attains a cube-root rate driven by the intrinsic entropy of the monotone transport map class. When distributions are observed through empirical samples, an additional approximation error enters, and our analysis characterizes the phase transition determined by the relative growth rates of the number of subjects and the within-distribution sample size.

From a computational perspective, the proposed formulation leads to an efficient and scalable estimation procedure. Leveraging the explicit quantile representation of one-dimensional Wasserstein distances, the infinite-dimensional optimization problem reduces to a sequence of finite-dimensional convex programs. Specifically, updating each transport map amounts to a weighted isotonic regression, while updating the weight vector corresponds to a constrained least-squares problem over the simplex. This block-wise structure yields stable numerical implementation and scales naturally with the number of predictors. 

Extensive simulation studies illustrate the finite-sample predictive behavior of the proposed method. In particular, we compare root mean squared prediction errors (RMSEs) against existing OT \citep{ghodrati2022distribution} and GOT \citep{zhu2023geodesic} regression approaches across a range of scenarios, including both single- and multi-predictor settings. A real data application to mortality forecasting further illustrates how the proposed framework yields accurate predictions together with interpretable assessments of predictor-specific distributional effects.

The remainder of this article is organized as follows. We briefly review the mathematical foundations of the Wasserstein space in Section~\ref{sec:pre}, and then present our regression model in Section~\ref{sec:model}, including the crucial identifiability results and estimation framework. In Section~\ref{sec:theory}, we establish the theoretical properties of the estimators. The computational algorithm is detailed in Section~\ref{sec:compute}. The proposed model is then validated through simulations in Section~\ref{sec:simu}, and Section~\ref{sec:real} illustrates the method through an application to mortality data. Proofs and auxiliary results are provided in the Supplementary Material.

\section{Preliminaries}\label{sec:pre}
\subsection{Optimal Transport}\label{sec:OT}
We first briefly review the Wasserstein space and optimal transport map. For more background, see, e.g., \citet{villani2021topics}, \citet{santambrogio2015optimal} and \citet{ambrosio2008gradient}. Let \(\mathcal{S} = [s_0,s_1]\) be a compact interval of \(\mathbb{R}\) and $\mathcal{W}_2(\mathcal{S})$ be the collection of probability measures on $\mathcal{S}$ with finite second-order moments, that is, $\mathcal{W}_2(\mathcal{S})=\left\{\mu \in \mathcal{P}(\mathcal{S}): \int_{\mathcal{S}}|x|^2 d \mu(x)<\infty\right\}$, where $\mathcal{P}(\mathcal{S})$ is the set of probability measures on $\mathcal{S}$. For two measures $\mu, \nu \in \mathcal{W}_2(\mathcal{S})$ that are absolutely continuous with respect to the Lebesgue measure, the squared $2$-Wasserstein distance admits the representation
\[
d_{\mathcal{W}}^2(\mu, \nu)
= \int_{\mathcal{S}} \{ F_\nu^{-1} \circ F_\mu(x) - x \}^2 \, \mathrm{d}\mu(x)
= \int_0^1 \left| F_\mu^{-1}(p) - F_\nu^{-1}(p) \right|^2 \, \mathrm{d}p,
\]
where $F^{-1}_{\nu}$ is the quantile function of $\nu$, and $F_\mu$ is the cumulative distribution function of $\mu$. The corresponding optimal transport map from $\mu$ to $\nu$ is given by
\(
T_{\mu\to\nu} := F_\nu^{-1} \circ F_\mu,
\)
which is a monotone map pushing $\mu$ forward to $\nu$. 
Motivated by this explicit representation, we will work extensively with transport maps. 

\subsection{Wasserstein Fréchet Mean}
We next recall the notion of the Fréchet mean in the Wasserstein space \citep{agueh2011barycenters}, which is defined through the metric structure of $\mathcal{W}_2(\mathcal{S})$. At the population level, let $\xi$ be a random probability measure in $\mathcal{W}_2(\mathcal{S})$ with distribution $P$. A Wasserstein Fréchet mean of $\xi$ is defined as any minimizer of the functional
\[
F(b) = \int_{\mathcal{W}_2(\mathcal{S})} d_{\mathcal{W}}^2(b, \xi) \mathrm{d} P(\xi), \, b \in \mathcal{W}_2(\mathcal{S}).
\]
In the finite-sample setting, let $\mu_1,\ldots,\mu_n \in \mathcal{W}_2(\mathcal{S})$ and let $\lambda_1,\ldots,\lambda_n$ be nonnegative weights satisfying $\sum_{i=1}^n \lambda_i = 1$. 
The corresponding (weighted) Wasserstein Fréchet mean is defined as any minimizer of
\[
F(b) = \sum_{i=1}^n \lambda_i \, d_{\mathcal{W}}^2(b,\mu_i), \, b \in \mathcal{W}_2(\mathcal{S}).
\]
Existence of such a minimizer is guaranteed and uniqueness holds if at least one $\mu_i$ is absolutely continuous with respect to the Lebesgue measure. Under the one-dimensional optimal transport setting of Section~\ref{sec:OT}, the Wasserstein Fréchet mean admits a simple characterization in terms of quantile functions. 
Specifically, if $\mu_1,\ldots,\mu_n$ have quantile functions $F_{\mu_1}^{-1},\ldots,F_{\mu_n}^{-1}$, then the Fréchet mean $\bar{\mu}$ has quantile function
\[
F_{\bar{\mu}}^{-1}(p) = \sum_{i=1}^n \lambda_i F_{\mu_i}^{-1}(p), \quad p\in(0,1).
\]
An analogous characterization holds at the population level by replacing the finite sum with an expectation.

\subsection{Notation}
We conclude this section by introducing some notational conventions. For functions $f$ and $g$, we write $f \circ g$ for their composition, defined by
$(f \circ g)(x) = f\{g(x)\}$.
For $a,b \ge 0$, we write $a \lesssim b$ or equivalently $b \gtrsim a$ 
if there exists a universal constant $C>0$ such that $a \le C b$.
The support of a probability measure $P$ is denoted by $\spt{P}$.
For a measurable function $f:\mathcal{S}\to\mathbb{R}$ and a probability measure $\mu$ on $\mathcal{S}$,
we write $\|f\|_{L^2(\mu)}$ for its $L^2$ norm with respect to $\mu$.
When $\mu$ is the Lebesgue measure on $[0,1]$, we simply write $\|f\|_{L^2}$. Let $\Delta^p$ denote the probability simplex in $\mathbb R^{p+1}$,
\[
\Delta^p := \Bigl\{ \bm{\alpha} = (\alpha_0,\alpha_1,\ldots,\alpha_p)^\top \in [0,1]^{p+1} :
\sum_{j=0}^p \alpha_j = 1 \Bigr\}.
\]
For a measurable map $T:\mathcal S\to\mathcal S$ and a probability measure $\mu$ on $\mathcal S$, 
the push forward measure $T\#\mu$ is defined by
\(
(T\#\mu)(A)=\mu(T^{-1}(A)),\, A\subset\mathcal S\ \text{measurable}.
\)

\section{Methodology}\label{sec:model}
We now formally introduce our regression model designed for scenarios with multiple predictor distributions. 
Let $(\bm{\xi},\eta) = (\xi_1,\xi_2,\ldots,\xi_p,\eta)$ be random elements taking values in $(\mathcal{W}_2(\mathcal{S}))^p \times \mathcal{W}_2(\mathcal{S})$ with joint distribution $P$, and assume that each distribution is absolutely continuous with respect to the Lebesgue measure on $\mathcal{S}$. In analogy with mean regression models in Euclidean space, where the regression function is the conditional expectation \(f(X) := \mathbb{E}(Y\mid X)\), we define a conditional Fréchet mean type regression operator 
\(\Gamma : (\mathcal{W}_2(\mathcal{S}))^p \to \mathcal{W}_2(\mathcal{S})\) as \[\Gamma(\bm{\xi}) := \mathop{\arg\min}_{b\in \mathcal{W}_2(\mathcal{S})} \int_{\mathcal{W}_2(\mathcal{S})} d_{\mathcal{W}}^2(b, \eta) \, \mathrm{d}P(\eta \mid \bm{\xi}),\]
where the minimizer is unique under mild regularity conditions.

To obtain a tractable and interpretable regression structure, we introduce a fixed and known reference distribution $\xi_0$, which plays a role analogous to an intercept. We assume that the regression operator $\Gamma(\bm{\xi})$ can be expressed as a weighted Wasserstein Fréchet mean of suitably transformed predictor distributions:
\begin{equation}\label{eq:reg_operator}
\Gamma(\bm{\xi}) = \argmin_{b\in\mathcal{W}_2(\mathcal{S})} \sum_{j=0}^{p} \alpha_j d_{\mathcal{W}}^2(b,T_j\#\xi_j) ,\quad \bm{\xi}\in(\mathcal{W}_2(\mathcal{S}))^p,
\end{equation}
where $\bm{\alpha}=(\alpha_0,\alpha_1,\ldots,\alpha_p)\in \Delta^{p}$ is a vector of nonnegative weights summing to one, $\bm{T}=(T_0,T_1,\ldots,T_p)\in\mathcal{T}^{p+1}$, and $\mathcal{T}$ denotes the collection of admissible optimal transport maps on $\mathcal{S}$,
\[ \mathcal{T} := \{ T \in \mathcal{S}\to\mathcal{S} : T \text{ is increasing},\, T(s_0)=s_0,\, T(s_1)=s_1 \}.\]
  
In the proposed model, each $T_j$ describes a shape deformation of the predictor distribution $\xi_j$, and the scalar weight $\alpha_j$ determines its relative contribution. Taken together, these two components form the ``regression coefficient'' of the predictor distribution \( \xi_j \). 
To anchor the model, a fixed and known reference distribution $\xi_0$ is introduced and can be chosen arbitrarily, provided that it is absolutely continuous with respect to the Lebesgue measure on $\mathcal S$. Although a different choice of $\xi_0$ alters the corresponding optimal transport map $T_0$, the induced term $T_0\#\xi_0$ remains unchanged. Thus, the reference distribution plays a role analogous to an intercept while leaving the overall regression operator invariant. In practice, common choices for the reference distribution include the uniform distribution on $\mathcal S$ or the Fréchet mean of the response distributions.

Then, we define the regression model 
\begin{equation}\label{eq:model}
\eta = T_\varepsilon \# \Gamma^*(\bm{\xi}),\quad
\Gamma^*(\bm{\xi}) = \argmin_{b\in\mathcal{W}_2(\mathcal{S})} \sum_{j=0}^{p} \alpha_j^* d_{\mathcal{W}}^2(b,T^*_j\#\xi_j)
\end{equation}
for some true parameters $\bm{\alpha}^*\in \Delta^p$ and $\bm{T}^* \in \mathcal{T}^{p+1}$. The operator \( T_{\varepsilon} \) acts as a disturbance operator and satisfies the point-wise expectation \( \mathbb{E}(T_{\varepsilon}\mid\bm{\xi}) = \operatorname{id}_{\mathcal{S}} \), where $\operatorname{id}_{\mathcal{S}} : \mathcal{S} \to \mathcal{S}$ denotes the identity map. This condition is similar to the classical assumption that the expectation of the noise error is zero.

\begin{remark}
The structure in \eqref{eq:model} can be viewed as an analogue of multivariate linear regression
\[Y = \beta_0\cdot 1 + \sum_{j=1}^p\beta_j X_j + \epsilon,\] 
for $(X_1,\ldots,X_p,Y)\in \mathbb{R}^p\times\mathbb{R}.$ Each regression coefficient $\beta_j$ of $X_j$ can be decomposed into a direction term, $\beta_j/\abs{\beta_j}$ and a magnitude term, $\abs{\beta_j}$. In our framework, these two roles are played respectively by the transport map $T_j$ and the weight $\alpha_j$. An additional point of analogy is that the constant $1$ appearing as the intercept in the linear model may, in principle, be replaced by any fixed constant, with the coefficients adjusting accordingly but leaving the fitted regression function unchanged.
\end{remark}

\subsection{Identifiability}
A fundamental requirement of any regression framework is that its population parameters be identifiable: distinct parameter values should induce distinct regression operators. We now show that, under mild regularity conditions, the proposed model is identifiable. We first characterize the regression operator explicitly.
\begin{proposition}\label{prop:mt}
For any $\bm{\xi}=(\xi_1,\ldots,\xi_p)$, the regression operator $\Gamma(\bm{\xi})$ defined in \eqref{eq:reg_operator} is given by 
\[ F_{\Gamma(\bm{\xi})}^{-1} = \sum_{j=0}^{p}\alpha_jT_j\circ F_{\xi_j}^{-1} .\]
\end{proposition}

Suppose that two sets of parameters \( (\bm{\alpha}, \bm{T})\) and \((\bm{\alpha}^{'}, \bm{T}^{'})\), with all weights strictly positive, produce the same regression operator for all admissible predictors drawn from the product distribution. The mutual independence structure of the predictor distributions is imposed solely for establishing identifiability of the parameters and is not required for model formulation or estimation. That is, 
for all $\left(\xi_1,\xi_2,\ldots,\xi_p\right)\sim P_{\xi_1}\otimes P_{\xi_2}\otimes\ldots\otimes P_{\xi_p}$, where $P_{\xi_j}$ denotes the marginal distribution induced on $\xi_j$,
\begin{equation}
\sum_{j=0}^{p}(\alpha_jT_j-\alpha_j^{'}T_j^{'})\circ F_{\xi_j}^{-1} =0.
\end{equation}

To guarantee uniqueness, we impose the following mild assumptions. 
\begin{assumption}[Nondegeneracy]\label{asm:nondegenerate}
For almost every $q \in (0,1)$, the set
\(R_j^q := \{ F_{\xi_j}^{-1}(q) : \xi_j \sim P_{\xi_j} \}\) contains more than one point for each \(j=1,2,\ldots,p \).
\end{assumption}
\begin{assumption}[Positive densities]\label{asm:positive}
The reference distribution \(\xi_0\) and the mean measures \(\mathbb{E}_{P_{\xi_j}}\xi_j\) possess densities that are strictly positive almost everywhere on $\mathcal{S}$.
\end{assumption}
\begin{assumption}[Continuity of transport maps]\label{asm:continuity}
Each transport map $T_j\in \mathcal{T}$ is continuous on $\mathcal{S}$, $j=0,1,\ldots,p$.
\end{assumption}

Assumption~\ref{asm:nondegenerate} ensures that each predictor distribution exhibits sufficient variability across samples, so that its quantile function is not degenerate at any fixed level. Assumption~\ref{asm:positive} guarantees absolute continuity and positivity of the relevant measures, which implies that the images of the quantile sets covers the domain $\mathcal{S}$. Assumption~\ref{asm:continuity} rules out discontinuous transport maps; in the one-dimensional setting, optimal transport maps between absolutely continuous measures are monotone and hence continuous almost everywhere (and continuous on $\mathcal{S}$ under mild regularity).

Under these conditions, the equality of the two regression operators forces each difference $\alpha_jT_j-\alpha_j^{'}T_j^{'}$ to be constant on $\mathcal{S}$. Because all transport maps in $\mathcal{T}$ share fixed boundary conditions, the only such constant is zero, leading to:
\[\alpha_j = \alpha_j^{'}, T_j-T_j^{\prime}=0, j=1,2,\ldots,p.\]
The model is therefore identifiable.
\begin{theorem}[Identifiability]\label{thm:ident}
Under Assumptions~\ref{asm:nondegenerate}, \ref{asm:positive} and \ref{asm:continuity}, the parameter pair $(\bm{\alpha},\bm{T})$ in \eqref{eq:model} is uniquely identified from the distribution of the data.
\end{theorem}

\subsection{Estimation}
As discussed earlier, the variation of $\bm{\xi}$ in the regression operator is jointly governed by the weights $\bm{\alpha}$ and the transport maps $\bm{T}$. For notational convenience and to streamline the analysis of asymptotic properties, we introduce the following parameter space:
\[ \Theta = \{\bm{\alpha}\odot\bm{T} : \bm{\alpha}\in\Delta^{p},\, \bm{T}\in\mathcal{T}^{p+1}\} \subset (L^\infty(\mathcal{S}))^{p+1},\]
where $\odot$ denotes the point-wise product. Each $\bm{\theta}\in \Theta$ uniquely determines the regression operator $\Gamma_{\bm{\theta}}$. %

To estimate $\bm{\theta}$, we adopt an M-estimation framework based on a least-squares-type Wasserstein loss. For any $(\bm{\xi},\eta)\in\spt{P}$, define the loss
\begin{equation}\label{eq:loss}
 m_{\bm{\theta}}(\bm{\xi},\eta) = d_{\mathcal{W}}^2(\eta,\Gamma_{\bm{\theta}}(\bm{\xi})) = \Big\lVert F_{\eta}^{-1} - \sum_{j=0}^{p}\alpha_jT_j\circ F_{\xi_j}^{-1} \Big\rVert_{L^2}^2.
 \end{equation}
Let $M(\bm{\theta}) = \mathbb{E}_{(\bm{\xi},\eta)\sim P}\, m_{\bm{\theta}}(\bm{\xi},\eta)$
denote the population risk, and let
\begin{equation}\label{eq:Mn}
M_n(\bm{\theta}) = \frac{1}{n}\sum_{i=1}^n m_{\bm{\theta}}(\bm{\xi}_i,\eta_i)
\end{equation}
denote its empirical counterpart based on $n$ i.i.d. samples $\{(\bm{\xi}_i,\eta_i)\}_{i=1}^n$.

In practice, the true distributions $(\bm{\xi},\eta)$ are rarely observed directly. Instead, one typically observes i.i.d. samples from each distribution and constructs empirical estimators $\bm{\xi}^m$ and $\eta^m$ based on a sample of size $m$. As stated by \citet{ghodrati2022distribution}, if each transport map $T\in\mathcal{T}$ is B-Lipschitz continuous, the Wasserstein error is stable under the push-forward operator, i.e.,
\[
d_{\mathcal{W}}^2(T\#\xi_j,T\#\xi_j^m) \lesssim\, d_{\mathcal{W}}^2(\xi_j,\xi_j^m),
\]
As a result, the plug-in loss function $m_{\bm{\theta}}(\bm{\xi}^m,\eta^m)$ inherits the convergence rate of the empirical distributions. Define $M_n^m(\bm{\theta}) = \frac{1}{n}\sum_{i=1}^n m_{\bm{\theta}}(\bm{\xi}_i^m,\eta_i^m)$, we have the following proposition. 
\begin{proposition}\label{prop:measure_error}
    Let $(\xi^m_1,\ldots,\xi^m_p,\eta^m)$ be empirical estimates that satisfy
    \[
    d_{\mathcal{W}}(\xi^m_j,\xi_j)=\mathcal{O}_{\Pr}(r_m), d_{\mathcal{W}}(\eta^m,\eta)=\mathcal{O}_{\Pr}(r_m),
    \]
    for some sequence $r_m\searrow
    0$. If each $T_j \in \mathcal{T}$ is B-Lipschitz continuous, then 
    \[
    \sup_{\bm{\theta}\in\Theta}
\abs{M_n^m(\bm{\theta})-M_n(\bm{\theta})}
= \mathcal{O}_{\Pr}(r_m).
    \]
\end{proposition}

\begin{remark}
The exact form of the rate $r_m$ depends on the choice of the empirical estimators and the regularity assumptions imposed on the underlying distributions.
Under appropriate conditions, minimax-optimal rates can be established over structured classes of distributions using existing approaches in the literature \citep{2016Functional,Victor2016point}.
\end{remark}

We consider an estimator $\hat{\bm{\theta}}$ for $\bm{\theta}^*$ satisfying that 
\begin{equation}\label{eq:est}
M_n(\hat{\bm{\theta}}) - \inf_{\bm{\theta}\in\Theta} M_n(\bm{\theta}) = \mathcal{O}_{\Pr}(e_n) 
\end{equation}
for some deterministic $e_n \searrow 0$. 
Proposition~\ref{prop:measure_error} shows that the discrepancy between $M_n^m$ and $M_n$ is asymptotically negligible. Consequently, the same approximate optimality condition \eqref{eq:est} holds when the estimator is obtained from the empirical distributions, that is,
\[\hat{\bm{\theta}}=\argmin_{\bm{\theta}\in\Theta}M_n^m(\bm{\theta}),\]
and $e_n = r_{m_n}\searrow 0$ as $n\to\infty$. 
Thus, the asymptotic properties established for the idealized setting carry over directly to the practically relevant case in which distributions are estimated from samples.

\section{Statistical Analysis}\label{sec:theory}
In this section, we investigate the asymptotic properties of the estimator constructed through \eqref{eq:est}. Our analysis proceeds by introducing a predictive semi-norm that naturally characterizes the geometry of the regression operator and enables a quadratic expansion of the population risk. For $\bm{f},\bm{g} \in (L^\infty(\mathcal{S}))^{p+1}$, define 
\[ \inprod{\bm{f}}{\bm{g}}_P = \E_{\bm{\xi}\sim P_{\bm\xi}}\, \bigg\{ \Big\langle \sum_{j=0}^{p}f_j\circ F_{\xi_j}^{-1}, \sum_{j=0}^{p}g_j\circ F_{\xi_j}^{-1} \Big\rangle_{L^2} \bigg\} \quad\text{and}\quad 
\norm{\bm{f}}_P = \inprod{\bm{f}}{\bm{f}}_P^{1/2} ,\]
where $\langle\cdot,\cdot\rangle_{L^2}$ denotes the usual inner product on $L^2(0,1)$. The above construction defines a semi-inner product and a semi-norm on $(L^\infty(\mathcal{S}))^{p+1}$; see Supplementary Material S3.4 for formal verification. Intuitively, the predictive semi-norm $\|\bm{f}\|_P$ measures the $L^2$ discrepancy between the induced prediction maps generated by $\bm{f}$, averaged over the distribution of the predictors.

The predictive semi-norm induces a natural metric on the parameter space $\Theta$. Define the diameter of the parameter space as $\operatorname{diam}\Theta = \sup_{\bm{\theta},\bm{\theta}'\in\Theta} \norm{\bm{\theta}-\bm{\theta}'}_P$. Since the outputs of all regression operators lie in $\mathcal{W}_2(\mathcal{S})$ with support contained in $[s_0,s_1]$, we obtain the global bound \((\operatorname{diam}\Theta)^2 \le (s_1-s_0)^2.\) This global bound characterizes the overall scale of the parameter space $\Theta$. 

The next lemma shows that the population risk is globally quadratic around $\bm{\theta}^*$ under the predictive semi-norm, which plays the role of a strong identification condition.
\begin{lemma}[Quadratic Growth]\label{lem:est-iden}
For every $\bm{\theta}\in\Theta$,
\[M(\bm{\theta})-M(\bm{\theta}^*) = \norm{\bm{\theta}-\bm{\theta}^*}_P^2 .\]
\end{lemma}
This identity reveals that the predictive semi-norm coincides with the curvature of the population risk, thereby providing an intrinsic metric for analyzing the convergence behavior of the estimator. Building on this deterministic structure, we next impose a continuity condition to control the stochastic fluctuations of the empirical process over local neighborhoods of $\Theta$. 
\begin{assumption}[Continuity]\label{asm:cts}
There exists a constant $\Lambda>0$ such that for all $\bm{f}\in(\Theta-\Theta)$ and all $(\bm{\xi},\eta)\in\spt{P}$,
\[ \Big\lVert \sum_{j=0}^{p}f_j\circ F_{\xi_j}^{-1} \Big\rVert_{L^2} \le \Lambda\norm{\bm{f}}_P,\]
where $(\Theta-\Theta):= \{ \bm{f} - \bm{g} : \bm{f}, \bm{g} \in \Theta \}
\subset (\mathrm{BV}(\mathcal{S}))^{p+1}$, and $\mathrm{BV}(\mathcal{S})$
denotes the space of real-valued functions on $\mathcal{S}$ with bounded variation, i.e., \(\int_{s_0}^{s_1} | \mathrm{d} f(s) | < \infty\).
\end{assumption}

\begin{remark}\label{rmk:cts}
Assumption~\ref{asm:cts} is used to control the local entropy numbers of $\Theta$. In the single-component setting, this continuity condition is implied by requiring that the random measure $\xi$ has a bounded density with respect to $\E\xi$. This requirement is closely related to the regularity condition considered in \citet{ghodrati2022distribution}.
\end{remark}

\begin{theorem}\label{thm:rate}
Under Assumption~\ref{asm:cts}, the estimator $\hat{\bm{\theta}}$ in \eqref{eq:est} satisfies that 
\[ \lVert\hat{\bm{\theta}}-\bm{\theta}^*\rVert_P = \mathcal{O}_{\Pr}(e_n^{1/2}+n^{-1/3}) .\]
\end{theorem}
The proof combines the quadratic growth property in
Lemma~\ref{lem:est-iden} with the local empirical process bound established in Supplementary Material S3.7.

\begin{remark}
Theorem~\ref{thm:rate} implies prediction consistency under the predictive semi-norm, in the sense that
$\|\hat{\bm{\theta}}-\bm{\theta}^*\|_P \to 0$ in probability provided that $e_n \to 0$.
Notably, the convergence rate is established with respect to the predictive semi-norm, which directly quantifies prediction accuracy rather than componentwise parameter estimation. Componentwise consistency of the weight vector $\bm{\alpha}$ and transport maps $\bm{T}$ can be obtained under additional regularity conditions on the parameterization, but is not required for prediction consistency and is therefore not pursued here.

The $n^{-1/3}$ term reflects the intrinsic non-Euclidean geometry induced by transport maps in the Wasserstein space.
In particular, the cube-root rate is driven by the entropy of the monotone transport map class, while the simplex constraint on the weights contributes only finite-dimensional complexity and does not affect the leading order.
When the regression operator is constructed using empirical distributions based on samples of size $m=m_n$, the additional approximation error enters through $e_n=r_{m_n}$, leading to a phase transition determined by the relative growth rates of $n$ and $m$.
This reveals a two-stage error structure, where the statistical error governed by $n^{-1/3}$ competes with the sampling-induced approximation error from estimating the input distributions.
In particular, when $m_n$ grows sufficiently fast so that $e_n^{1/2} = o(n^{-1/3})$, the statistical error dominates and the estimator attains the intrinsic $n^{-1/3}$ rate, whereas slower growth of $m_n$ results in the sampling-induced error dominating the overall performance.
\end{remark}

\section{Implementation and Computation Details}\label{sec:compute}
With the theoretical framework established, we next describe the computational procedure required to implement the proposed method. In particular, we compute the estimator of $\bm{\theta}$ using a block coordinate descent scheme, which alternates between updating the weight vector $\bm{\alpha}$ and the collection of transport maps $\bm{T}$. 

When estimating $\bm{T}$, we adopt a back-fitting-type scheme: each update of $T_k$ is performed while treating the remaining components $\{T_j, j\neq k\}$ as fixed. A central challenge lies in approximating and optimizing the Fréchet sum of squares in the Wasserstein space. To facilitate computation, we approximate the $L^2$ loss \eqref{eq:Mn} by a Riemann sum over a user-defined partition of $\mathcal{S}$. 
Using a change of variables, the objective can be rewritten in terms of optimal transport maps, and subsequently discretized as
\[
\frac{1}{n}\sum_{i=1}^n\sum_{r=1}^t \left|T_{\xi_{ki}\to \eta_i}(x_r) - \alpha_k T_k(x_r) - \sum_{j\neq k}\alpha_j T_j\circ T_{\xi_{ji}\to\xi_{ki}}(x_r)\right|^2\xi_{ki}(h_r),
\]
where \(t\) denotes the number of user-defined nodes \(\{x_r\}^t_{r=1}\) in an interval partition \(\{I_r\}^t_{r=1}\) of \(\mathcal{S}\), \(h_r = |I_r|\). These discrete approximations allow us to reduce the infinite-dimensional optimization problem to a finite sum over nodes, thereby enabling efficient numerical implementation. 
We then adopt an iterative back-fitting procedure. At iteration $\ell$, with current estimates $\bm{\alpha}^{(\ell-1)}$ in the interior of the simplex $\Delta^p$ and all other components $\{T_j^{(\ell)}: j<k\}$ and $\{T_j^{(\ell-1)}: j>k\}$ fixed, the $k$th transport map is updated by solving
\begin{equation}\label{T1}
\begin{aligned}
T_k^{(\ell)} = 
\argmin_{T_k\in\mathcal{T}} \frac{1}{n}\sum_{i=1}^n\sum_{r=1}^t \Big| &T_{\xi_{ki}\to \eta_i}(x_r) - \sum_{j<k}\alpha_j^{(\ell-1)} T_j^{(\ell)}\circ T_{\xi_{ji}\to\xi_{ki}}(x_r)\\
&- \alpha_k^{(\ell-1)} T_k(x_r) - \sum_{j>k}\alpha_j^{(\ell-1)} T_j^{(\ell-1)}\circ T_{\xi_{ji}\to\xi_{ki}}(x_r)\Big|^2\,\xi_{ki}(h_r).\\
\end{aligned}
\end{equation}
By defining $w_{i r}=\xi_{ki}(h_r)$, $z_r=T_k(x_r)$ and  
\begin{equation*}
y_{i r} = \left(T_{\xi_{ki}\to \eta_i}(x_r) - \sum_{j<k}\alpha_j^{(\ell-1)} T_j^{(\ell)}\circ T_{\xi_{ji}\to\xi_{ki}}(x_r) - \sum_{j>k}\alpha_j^{(\ell-1)} T_j^{(\ell-1)}\circ T_{\xi_{ji}\to\xi_{ki}}(x_r)\right)/\alpha_k^{(\ell-1)}, 
\end{equation*}
the approximate minimization \eqref{T1} reduces to the following problem:
$$
\begin{aligned}
& \min f(z)=\frac{1}{n}\sum_{i=1}^n \sum_{r=1}^t \left(w_{i r} \left|y_{i r}-z_r\right|^2 \right)  \\
& \text { subject to } s_0 = z_1 < z_2 < \cdots < z_t = s_1.
\end{aligned}
$$
Setting $\begin{gathered}\hat{y}_r=\frac{\sum_{i=1}^n w_{i r} y_{i r}}{\sum_{i=1}^n w_{i r}} \text{ and } \tilde{w}_r=\frac{1}{n}\sum_{i=1}^n w_{i r},\end{gathered}$
we can rewrite
$$\frac{1}{n}\sum_{i=1}^n w_{i r}\left(y_{i r}-z_r\right)^2=\frac{1}{n}\sum_{i=1}^n w_{i r}\left(y_{i r}-\hat{y}_r\right)^2+\tilde{w}_r\left(z_r-\hat{y}_r\right)^2.$$
Thus, the problem is equivalent to a weighted isotonic regression:
\begin{equation}\label{eq:isotonic}
\begin{aligned}
& \min f(z)=\sum_{r=1}^t \tilde{w}_r \left(z_r - \hat{y}_r\right)^2 \\ 
& \text { subject to } s_0 = z_1 < z_2 < \cdots < z_t = s_1 .
\end{aligned}
\end{equation}

Consequently, all parameters are estimated through an alternating iterative procedure, summarized in Algorithm~\ref{algorithm-1}, which alternates between:
(i) updating each transport map $T_k$ via a backfitting step that solves a weighted isotonic regression problem, and 
(ii) updating the weight vector $\bm{\alpha}$ by solving a constrained least-squares problem over the simplex.

\begin{algorithm}[!ht]
\caption{Alternating Iterative Procedure for the proposed Model}\label{algorithm-1}
\begin{algorithmic}[1] 	
\Require $\{\bm{\xi}_i,\eta_i\}, i=1,\ldots,n$, reference distribution $\xi_0$, grid size $t$, initial maps $T_j^{(0)}=\operatorname{id}_{\mathcal{S}},\, j=0,1,\ldots,p.$\\
\(
\bm{\alpha}^{(0)} = \argmin_{\bm{\alpha}\in\Delta^p} \frac{1}{n}\sum_{i=1}^n\Big\lVert F_{\eta_i}^{-1} - \sum_{j=0}^{p}\alpha_jT_j^{(0)}\circ F_{\xi_{ji}}^{-1} \Big\rVert_{L^2}^2.
\)
\For{$\ell=1,2,\ldots$}
        \For{$k=0,1,\ldots,p$}
    \State Update $T_k^{(\ell)}$ by solving the subproblem \eqref{T1},
        with $\{T_j^{(\ell)}: j<k\}$ and $\{T_j^{(\ell-1)}: j>k\}$ held fixed.
    \EndFor
    \State
    \(
\bm{\alpha}^{(\ell)} = \argmin_{\bm{\alpha}\in\Delta^p} \frac{1}{n}\sum_{i=1}^n\Big\lVert F_{\eta_i}^{-1} - \sum_{j=0}^{p}\alpha_jT_j^{(\ell)}\circ F_{\xi_{ji}}^{-1} \Big\rVert_{L^2}^2.\)
	\If{convergence}
	\State Stop and denote $\{\widehat{\alpha}_j,\widehat{T}_j\}_{j=0,1,\ldots,p}=\{\alpha_j^{(\ell)},T_j^{(\ell)}\}_{j=0,1,\ldots,p}$;
        \EndIf
	\EndFor
	\Ensure 
 \(\widehat{\bm{\alpha}}, \widehat{\bm{T}}, \widehat{\bm{\theta}}=\widehat{\bm{\alpha}}\odot\widehat{\bm{T}}.\)
\end{algorithmic} 
\end{algorithm} 

In matrix form, let \(\mathbf{z}=(z_1,\ldots,z_t)^{\top}\), \(\mathbf{y}=(\hat{y}_1,\ldots,\hat{y}_t)^{\top}\), and \(W=\operatorname{diag}(\tilde{w}_1,\ldots,\tilde{w}_t)\). The problem \eqref{eq:isotonic} becomes the standard quadratic program with linear inequality constraints:
\[\min_{\mathbf{z} \in \mathbb{R}^t}  (\mathbf{z}-\mathbf{y})^{\top} W(\mathbf{z}-\mathbf{y}), \text{ s.t. }s_0 = z_1 < z_2 < \cdots < z_t = s_1. \]
This procedure reduces the original functional optimization problem into a sequence of finite-dimensional quadratic programs, which can be solved efficiently using standard convex optimization tools.

\section{Simulations}\label{sec:simu}
We present a set of extensive simulation studies designed to assess the proposed model under both single and multiple predictors settings. In addition, we compare its performance with that of the OT model \citep{ghodrati2022distribution} and the GOT model \citep{zhu2023geodesic}.
\subsection{Single-predictor}\label{sec:sigle_simu}
Although our model is developed for multiple predictors ($p\geq 1$), the case $p=1$ represents a notable special case. We therefore begin the simulation study by first presenting results for this single-predictor setting.

We choose the uniform distribution on \(\mathcal{S} = [0,1]\) as the reference probability measure \(\xi_0\). For the random predictor distributions \(\{\xi_{1i}\}_{i=1}^{n}\), we consider Beta distributions whose parameters are uniformly distributed random variables on the interval \([1,5]\), i.e.,
\[
f_{\xi_{1i}}(x) = b_{a_i, b_i}(x),\quad a_i \sim \mathrm{U}[1,5],\quad b_i \sim \mathrm{U}[1,5],\quad i = 1, \cdots, n.
\]
For noise mappings \(T_{\epsilon_i}\), we adopt the class of random optimal maps proposed by \citet{Victor2016point}. Let \(k\) be an integer; we define the mapping \(g_k: [0,1] \to [0,1]\) as follows:
\begin{equation}\label{noise}
g_0(x) = x,\quad g_k(x) = x - \frac{\sin(\pi kx)}{|k|\pi}, \quad k \in \mathbb{Z} \setminus \{0\}.
\end{equation}
These functions increase strictly and are smooth, satisfying \(g_k(0) = 0\) and \(g_k(1) = 1\) for any \(k\). Replacing \(k\) with an integer-valued random variable \(K\) whose distribution is symmetric about zero, it follows that for all \(x \in [0, 1]\), \(E[g_K(x)] = x\), as required by the model.

Regarding the optimal maps that constitute the regression operators, \(T_0^*\) and \(T_1^*\), we set
\[
T_0^* = g_4,\quad T_1^* = g_3.
\]
After generating random \(\xi_{1i}\) and \(T_{\epsilon_i}\), the response distribution can be obtained according to different values of \(\alpha_1^* \in \{0, 0.25, 0.5, 0.75, 1\}\), that is, \(\eta_i = T_{\epsilon_i}\#\Gamma(\xi_{1i})\),
\begin{equation}\label{generate_single}
F_{\eta_i}^{-1}(\cdot) = g_{K_i}\circ\left\{(1-\alpha_1^*) T_0^*\circ F_{\xi_{0}}^{-1}(\cdot) + \alpha_1^*T_1^*\circ F_{\xi_{1i}}^{-1}(\cdot)\right\}.
\end{equation}
Since we do not directly observe these quantile functions $F_{\eta_i}^{-1}, F_{\xi_{1i}}^{-1}$ in practice, we assume that we have the following random sample observations $\{F_{ \xi_{1i}}^{-1}\left(u_{i 1}\right), F_{ \xi_{1i}}^{-1}\left(u_{i 2}\right), \ldots, F_{ \xi_{1i}}^{-1}\left(u_{i m}\right)\}$ and $\{F_{\eta_i}^{-1}\left(v_{i 1}\right), F_{\eta_i}^{-1}\left(v_{i 2}\right)$ $,\ldots,$ $F_{\eta_i}^{-1}\left(v_{i m}\right)\}$, where $u_i, v_i$ s are independently generated from the distribution $\mathrm{U}(0,1)$. Based on observations, the quantile functions \(F_{\xi_{1i}}^{-1}(q)\) and \(F_{\eta_i}^{-1}(q)\) are estimated based on empirical quantiles on a grid of \(q\in [0,1]\). 

The table below display the estimation results for different parameter settings based on \(m\) random sample observation points and \(n\) groups of random probability distributions. We evaluated the performance in terms of $\|\widehat{\bm{\theta}}^j-\bm{\theta}^*\|_P$, $\left|\widehat{\alpha}_1^j-\alpha_1^*\right|$, $\|\widehat{T}_0^j-T_0^*\|_{L_2}$ and $\|\widehat{T}_1^j-T_1^*\|_{L_2}$. Here, $\widehat{\bm{\theta}}^j$, $\widehat{\alpha}_1^j, \widehat{T}_0^j, \widehat{T}_1^j$ are the estimates from the $j$th replicated dataset. Table~\ref{single_simu} reports the mean and standard deviation results, where $\|\widehat{\bm{\theta}}-\bm{\theta}^*\|_P$ is computed on a test set with a sample size of $0.3n$.

\begin{table}[!htbp]
\centering
\renewcommand{\arraystretch}{0.7}
\caption{Monte Carlo mean (standard deviation) based on 100 replications in single-predictor setting.}
\label{single_simu}
\begin{tabular}{@{}ccccccc@{}}
\toprule\noalign{}
$\alpha_1^*$  & $n$&
$m$ & $\left\|\widehat{\bm{\theta}}-\bm{\theta}^*\right\|_P$ & $\left|\widehat{\alpha}_1-\alpha_1^*\right|$ & $\left\|\widehat{T}_0-T_0^*\right\|_{L^2}$ & $\left\|\widehat{T}_1-T_1^*\right\|_{L^2}$\\
\bottomrule\noalign{}
\multirow{6}{*}{0}& \multirow{2}{*}{50}&200&0.017(0.005) &0.036(0.023)&
                0.016(0.006) &\diagbox{}{}\\
&&400&0.014(0.005) &0.035(0.020)&
                0.012(0.006) &\diagbox{}{}\\
                \cline{2-7}
& \multirow{2}{*}{200}&200&0.012(0.002)&0.030(0.013)&
                0.012(0.003) &\diagbox{}{}\\
&&400&0.008(0.002)& 0.018(0.010)&
                0.008(0.003)&\diagbox{}{} \\
                \cline{2-7}
& \multirow{2}{*}{400}&200&0.011(0.001) &0.025(0.009)&
                0.011(0.002) &\diagbox{}{}\\
&&400&0.007(0.001) &0.017(0.008)&
                0.007(0.002) &\diagbox{}{}\\
                \midrule\noalign{}
\midrule\noalign{}
\multirow{6}{*}{0.25}& \multirow{2}{*}{50}&200&0.018(0.005) &0.028(0.020)&
                0.021(0.009) &0.056(0.025)\\
&&400&0.015(0.005) &0.024(0.021)&
                0.016(0.007) &0.043(0.017)\\
                \cline{2-7}
& \multirow{2}{*}{200}&200&0.011(0.002)& 0.017(0.012)&
                0.016(0.005)& 0.035(0.011)\\
&&400&0.008(0.002) &0.012(0.009)&
                0.011(0.005) &0.029(0.011)\\
                \cline{2-7}
& \multirow{2}{*}{400}&200&0.009(0.001)& 0.012(0.008)&
                0.015(0.004) &0.033(0.011)\\
                
&&400& 0.006(0.001) &0.010(0.008)&
                0.009(0.003) &0.023(0.009) \\ 
\midrule\noalign{}
\midrule\noalign{}
\multirow{6}{*}{0.50}& \multirow{2}{*}{50}&200&
0.017(0.005)&0.027(0.020)&
0.023(0.008)&0.026(0.010)\\
&&400&
0.015(0.005)&0.023(0.022)&
0.021(0.010)&0.024(0.009)\\
\cline{2-7}

& \multirow{2}{*}{200}&200&
0.009(0.002)&0.015(0.011)&
0.014(0.003)&0.014(0.004)\\
&&400&
0.008(0.002)&0.014(0.011)&
0.012(0.004)&0.012(0.004)\\
\cline{2-7}

& \multirow{2}{*}{400}&200&
0.007(0.001)&0.010(0.007)&
0.012(0.003)&0.010(0.003)\\
&&400&
0.006(0.001)&0.010(0.009)&
0.009(0.002)&0.009(0.003)\\
\midrule\noalign{}
\midrule\noalign{}
\multirow{6}{*}{0.75}& \multirow{2}{*}{50}&200&
0.016(0.005)&0.038(0.026)&
0.060(0.025)&0.025(0.011)\\
&&400&
0.014(0.005)&0.027(0.022)&
0.049(0.024)&0.020(0.008)\\
\cline{2-7}

& \multirow{2}{*}{200}&200&
0.009(0.002)&0.029(0.018)&
0.062(0.033)&0.024(0.013)\\
&&400&
0.008(0.002)&0.023(0.014)&
0.042(0.019)&0.016(0.008)\\
\cline{2-7}

& \multirow{2}{*}{400}&200&
0.007(0.001)&0.026(0.018)&
0.058(0.030)&0.021(0.011)\\
&&400&
0.006(0.002)&0.019(0.012)&
0.041(0.019)&0.015(0.007)\\
\midrule\noalign{}
\midrule\noalign{}
\multirow{6}{*}{1}& \multirow{2}{*}{50}&200&
0.015(0.005)&0.062(0.041)&\diagbox{}{}
&0.028(0.015)\\
&&400&
0.013(0.005)&0.034(0.032)&\diagbox{}{}
&0.017(0.010)\\
\cline{2-7}

& \multirow{2}{*}{200}&200&
0.009(0.003)&0.051(0.033)&\diagbox{}{}
&0.025(0.015)\\
&&400&
0.008(0.002)&0.036(0.025)&\diagbox{}{}
&0.016(0.009)\\
\cline{2-7}

& \multirow{2}{*}{400}&200&
0.007(0.002)&0.048(0.031)&\diagbox{}{}
&0.023(0.014)\\
&&400&
0.006(0.002)&0.029(0.023)&\diagbox{}{}
&0.014(0.009)\\
\bottomrule\noalign{}
\end{tabular}
\end{table}

\begin{remark}\label{remark1}
It should be noted that when $\alpha_1^* = 0$ or $1$, the corresponding target distribution is determined solely by one of the predictor distributions. Consequently, $T_1$ is not identifiable when $\alpha_1^* = 0$ (similarly, $T_0$ is not identifiable when $\alpha_1^* = 1$). Therefore, the corresponding estimation errors are not reported in the table.
\end{remark}

Based on the discussion above, our model can be viewed as an extension of the OT model \citep{ghodrati2022distribution}. Specifically, in the single predictor setting, our model incorporates an additional reference distribution, leading to what we term the Multi-transport Distributional Regression (MTDR) model. To evaluate its performance, we compare the RMSEs (\(\sqrt{\frac{1}{0.3n}\sum_{k=n+1}^{1.3n}d_{\mathcal{W}}^2(\eta_k,\widehat{\eta}_k)}\) evaluated on a test set with a sample size of \(0.3n\)) of the MTDR and OT models under the same data-generating setting \eqref{generate_single} for various values of $\alpha_1^*$, with the results summarized in the first two columns of Table~\ref{rmse_single}. When $\alpha_1^* = 1$, the MTDR model reduces to the OT model
\[
\eta_i = T_{\epsilon_i} \# \left[T_1\#\xi_{1i}\right], \quad \{ \xi_{1i}, \eta_i \}_{i=1}^{n}.
\]
As shown in the first two columns of Table~\ref{rmse_single}, when $\alpha_1^* = 1$, the two yield identical results, which is expected. As $\alpha_1^*$ decreases, which corresponds to assigning a larger weight to the reference distribution, the regression effect of $\xi_1$ on $\eta$ becomes weaker. In this case, the OT model, which lacks the capacity to capture such a regression structure, yields noticeably larger RMSE values than the MTDR model, underscoring the clear performance improvement brought by incorporating the reference distribution.

\begin{table}[!htbp]
\centering
\caption{Comparison of RMSE values for different values of $\alpha_1^*$ under MTDR model setting \eqref{generate_single} \((n=m=200)\).
The smallest RMSE value in each row is highlighted in bold.}
    \label{rmse_single}
    \begin{tabular}{@{}cccc@{}}
\toprule\noalign{}
\(\alpha_1^*\)   &$RMSE_{MTDR}$&$RMSE_{OT}$&$RMSE_{GOT}$\\
\midrule\noalign{}
      1&\textbf{0.076(0.011)}& \textbf{0.076(0.011)}& 0.103(0.009)  
             \\
     0.7& \textbf{0.074(0.011)}& 0.096(0.010) &0.089(0.009)\\
      0.5 & \textbf{0.073(0.010)}& 
       0.125(0.009)& 
            0.082(0.009) \\ 
      0.3&\textbf{0.072(0.010)} &
       0.158(0.010) &
            0.076(0.009) \\
            0&\textbf{0.070(0.009)}& 0.213(0.011)& \textbf{0.070(0.009)}
            \\
            \bottomrule
            \end{tabular}
            \end{table}

Furthermore, we performed a comparative analysis against the GOT model proposed by \citet{zhu2023geodesic}. Specifically, the GOT model under a single predictor setting is formulated as
\[
\eta_i = \left[T_{\epsilon_i}\circ(\beta\odot T_{\bar\xi\to\xi_i})\right]\#\bar\eta,\quad \{ \xi_i, \eta_i \}_{i=1}^{n},
\]
or equivalently, from the perspective of quantile functions,
\[
F_{\eta_i}^{-1}(\cdot) = \left[T_{\epsilon_i}\circ(\beta\odot T_{\bar\xi\to\xi_i})\right]\circ F_{\bar\eta}^{-1}(\cdot),
\]
where $\bar\xi$ and $\bar\eta$ denote the Fréchet means of $\xi$ and $\eta$, respectively; $\beta \in \mathbb{R}$ is the true model parameter; $T_{\bar\xi\to\xi_{i}}$ is the optimal transport map from $\bar\xi$ to $\xi_{i}$; and $\{T_{\epsilon_i}\}_{i=1}^N$ are random perturbation maps. The definition of the operation $\beta \odot T_{\bar\xi \to \xi_i}$, together with additional details of the model specification, can be found in \citet{zhu2023geodesic}.

Under the same data-generating setting as in \eqref{generate_single}, the GOT results are reported in the third column of Table~\ref{rmse_single}. Notably, when $\alpha_1^* = 0$, the MTDR model degenerates to the case where the response distribution is independent of the predictor distribution and is generated solely by perturbing a fixed reference distribution.
\[
\eta_i = T_{\epsilon_i} \# \left[T_0\#\xi_0\right], \quad \{ \xi_i, \eta_i \}_{i=1}^{n}.
\]
This special case coincides with the \textit{pure intercept model} discussed in Remark 3 of \citet{ghodrati2022distribution}, i.e.
\[
T_0 = T_{\xi_0\to\bar\eta}=F^{-1}_{\bar\eta}\circ F_{\xi_0},
\]
which also corresponds to the GOT model with $\beta = 0$. Consequently, as shown in Table~\ref{rmse_single}, when $\alpha_1^* = 0$, the RMSE values of the MTDR and GOT models are identical. For other values of $\alpha_1^*$ and $\beta$, the different data-generating mechanisms lead to consistently higher RMSE values for the GOT model compared with the MTDR model in our simulation setting. A comparison of the three methods under the data-generating setting of the GOT model is provided in Supplementary Material S1.

\subsection{Multiple-predictor}
In the following section, we generalize to the multiple-predictor setting, starting with the case where $p=2$. The choice of \(\xi_0\) and \(T_{\epsilon_i}\) are the same as in Section~\ref{sec:sigle_simu}, the settings for $\xi_1$, $\xi_2$, $T_0^*$, $T_1^*$, and $T_2^*$ are as follows:
$$f_{\xi_{1i}}(x)=b_{a_{1i}, b_{1i}}(x), a_{1i} \sim \mathrm{U}[1,5], b_{1i} \sim \mathrm{U}[1,5],$$
$$f_{\xi_{2i}}(x) = b_{a_{2 i}, b_{2 i}}(x), a_{2i} \sim \mathrm{U}[2,6], b_{2i} \sim \mathrm{U}[2,6],$$
$$
T_0^* = g_4, \quad T_1^* = g_3, \quad T_2^* = g_{-5}.
$$
Then the response distribution can be obtained according to different values of \(\bm{\alpha}^{*}=(\alpha_0^*,\alpha_1^*,\alpha_2^*)^\top \in \{(0.3,0.35,0.35)^\top, (0.2,0.2,0.6)^\top,(0,0.5,0.5)^\top,(0,1,0)^\top\}\), that is, \(\eta_i = T_{\epsilon_i}\#\Gamma(\xi_{1i},\xi_{2i})\),
\begin{equation}\label{generate_multiple}
F_{\eta_i}^{-1}(\cdot) = g_{K_i}\circ\left\{\alpha_0^* T_0^*\circ F_{\xi_{0}}^{-1}(\cdot) + \alpha_1^*T_1^*\circ F_{\xi_{1i}}^{-1}(\cdot) + \alpha_2^*T_2^*\circ F_{\xi_{2i}}^{-1}(\cdot)\right\}.
\end{equation}

Similarly, we assume that we have the following random sample observations $\{F_{ \xi_{1i}}^{-1}\left(u_{i 1}\right),F_{ \xi_{1i}}^{-1}\left(u_{i 2}\right), \ldots, F_{ \xi_{1i}}^{-1}\left(u_{i n}\right)\}$, $\{F_{ \xi_{2i}}^{-1}\left(v_{i 1}\right),F_{ \xi_{2i}}^{-1}\left(v_{i 2}\right),\ldots,F_{ \xi_{2i}}^{-1}\left(v_{i n}\right)\}$ and $\{F_{\eta_i}^{-1}\left(w_{i 1}\right),$ $ F_{\eta_i}^{-1}\left(w_{i 2}\right), \ldots, F_{\eta_i}^{-1}\left(w_{i n}\right)\}$, where $u_i, v_i, w_i$ s are independently generated from the distribution $\mathrm{U}(0,1)$. Based on observations, the quantile functions \(F_{\xi_{1i}}^{-1}(q)\), \(F_{\xi_{2i}}^{-1}(q)\) and \(F_{\eta_i}^{-1}(q)\) are estimated based on empirical quantiles on a grid of $q$ values \(\in [0,1]\).

We also present the estimation results for different parameter settings based on \(m\) random sample observation points and \(n\) groups of random probability distributions. We evaluated the performance in terms of $\|\widehat{\bm{\theta}}-\bm{\theta}^*\|_P$, $\|\widehat{\bm{\alpha}}-\bm{\alpha}^*\|_{\ell^2}$, $\|\widehat{T}_0-T_0^*\|_{L_2}$, $\|\widehat{T}_1-T_1^*\|_{L_2}$ and $\|\widehat{T}_2-T_2^*\|_{L_2}$, as shown in Table~\ref{multi_simu1}.

\begin{table}[!htbp]
\centering
\renewcommand{\arraystretch}{0.85}
\caption{Monte Carlo mean (standard deviation) based on 100 replications in the multiple-predictor setting.}
    \label{multi_simu1}
\begin{tabular}{@{}cccccccc@{}}
\toprule\noalign{}
$\bm{\alpha}^*$  
& $n$&$m$ 
& $\left\|\widehat{\bm{\theta}}-\bm{\theta}^*\right\|_P$ & $\left\|\widehat{\bm{\alpha}}-\bm{\alpha}^*\right\|_{\ell^2}$ & $\left\|\widehat{T}_0-T_0^*\right\|_{L^2}$ & $\left\|\widehat{T}_1-T_1^*\right\|_{L^2}$
&$\left\|\widehat{T}_2-T_2^*\right\|_{L^2}$\\
\bottomrule\noalign{}
\multirow{6}{*}{$\left(
\begin{array}{c}
0.3\\
0.35\\
0.35
\end{array}
\right)$} & \multirow{2}{*}{50}&
200&0.020(0.006) &0.094(0.056) &0.057(0.032)& 0.050(0.024) &0.060(0.038)\\
&&400&0.019(0.005)& 0.082(0.049) &0.047(0.023)& 0.045(0.021) &0.049(0.026)\\
                \cline{2-8}
& \multirow{2}{*}{200}&
200&0.010(0.002) &0.043(0.022)& 0.028(0.011)& 0.024(0.009)& 0.026(0.009)\\
&&400&0.009(0.002) &0.041(0.021)& 0.026(0.010)& 0.022(0.009) &0.021(0.008)\\
                \cline{2-8}
& \multirow{2}{*}{400}&
200&0.008(0.002) &0.034(0.018)& 0.022(0.010)& 0.019(0.008) &0.021(0.010)\\
&&400&0.007(0.002) &0.032(0.018)& 0.018(0.008)& 0.015(0.006) &0.015(0.005)\\
                \midrule\noalign{}
\midrule\noalign{}
\multirow{6}{*}{$\left(
\begin{array}{c}
0.2\\
0.2\\
0.6
\end{array}
\right)$} & \multirow{2}{*}{50}&
200&0.020(0.005) & 0.096(0.051) & 0.084(0.047) & 0.077(0.037) & 0.038(0.013) \\
&&400&0.019(0.005) & 0.085(0.052) & 0.069(0.035) & 0.078(0.040) & 0.029(0.013) \\
\cline{2-8}
&\multirow{2}{*}{200}&
200&0.011(0.002) & 0.054(0.022) & 0.053(0.026) & 0.057(0.028) & 0.031(0.009) \\
&&400&0.009(0.002) & 0.044(0.024) & 0.041(0.016) & 0.043(0.020) & 0.020(0.007) \\
\cline{2-8}
&\multirow{2}{*}{400}&
200&0.008(0.002) & 0.046(0.018) & 0.051(0.022) & 0.052(0.024) & 0.032(0.009) \\
&&400&0.007(0.002) & 0.035(0.017) & 0.033(0.016) & 0.038(0.018) & 0.019(0.007) \\
\midrule\noalign{}
\midrule\noalign{}
\multirow{6}{*}{$\left(
\begin{array}{c}
0\\
0.5\\
0.5
\end{array}
\right)$} & \multirow{2}{*}{50}&
200&0.020(0.005) &0.113(0.061)&\diagbox{}{}& 0.050(0.027)& 0.048(0.027)\\
&&400&0.018(0.006) &0.089(0.055) &\diagbox{}{}& 0.039(0.025) &0.035(0.019)\\
\cline{2-8}
&\multirow{2}{*}{200}&
200&0.010(0.002)& 0.069(0.034) &\diagbox{}{}& 0.034(0.017) &0.024(0.010) \\
&&400&0.009(0.002) &0.052(0.027) &\diagbox{}{}& 0.021(0.010) &0.016(0.007) \\
\cline{2-8}
&\multirow{2}{*}{400}&
200&0.008(0.002)& 0.062(0.029) &\diagbox{}{}& 0.031(0.017) &0.020(0.009) \\
&&400&0.006(0.002)& 0.040(0.023) &\diagbox{}{}& 0.016(0.008) &0.012(0.006) \\
\midrule\noalign{}
\midrule\noalign{}
\multirow{6}{*}{$\left(
\begin{array}{c}
0\\
1\\
0
\end{array}
\right)$} & \multirow{2}{*}{50}&
200&0.017(0.005)& 0.123(0.056)&\diagbox{}{}& 0.040(0.019)  &\diagbox{}{}  \\
&&400&0.015(0.005) &0.085(0.048) &\diagbox{}{}& 0.026(0.014)&\diagbox{}{}  \\
\cline{2-8}
&\multirow{2}{*}{200}&
200&0.009(0.002) &0.105(0.042)&\diagbox{}{}& 0.040(0.017)  &\diagbox{}{} \\
&&400&0.008(0.002) &0.069(0.036) &\diagbox{}{}& 0.024(0.013) &\diagbox{}{}  \\
\cline{2-8}
&\multirow{2}{*}{400}&
200&0.008(0.002) &0.098(0.039) &\diagbox{}{}&0.039(0.016)&\diagbox{}{}  \\
&&400&0.006(0.002) &0.065(0.036) &\diagbox{}{}& 0.023(0.014) &\diagbox{}{}  \\
\bottomrule\noalign{}
\end{tabular}
\end{table}

\begin{remark}
In line with Remark~\ref{remark1}, when $\alpha_k^* = 0$, $k\in\{0,1,2\}$, the corresponding $T_k$ cannot be identified, as the target distribution depends on the other predictor distribution. Therefore, the estimation errors for $T_k$ are not reported in the table.
\end{remark}

Building on the previous comparison, we now turn to the multiple-predictor setting. Since the OT model cannot be directly applied in scenarios with more than one predictor distribution, we restrict our comparison to the GOT model. In this case, assume that there exists an unknown true ordering $j_1^*, j_2^*$ of indices $1,2$ of the two predictors that determines their order in the regression model, the GOT model with two predictors emerges as
\[
\eta_i = \left[T_{\epsilon_i}\circ(\beta_1\odot T_{\bar\xi_{j_1^*}\to\xi_{j_1^* i}})\circ (\beta_2\odot T_{\bar\xi_{j_2^*}\to\xi_{j_2^* i}})\right]\#\bar\eta,\quad \{ \xi_{1i},\xi_{2i}, \eta_i \}_{i=1}^{n},
\]
or equivalently, from the perspective of quantile functions,
\[
F_{\eta_i}^{-1}(\cdot) = \left[T_{\epsilon_i}\circ(\beta_1\odot T_{\bar\xi_{j_1^*}\to\xi_{j_1^*i}})\circ (\beta_2\odot T_{\bar\xi_{j_2^*}\to\xi_{j_2^* i}})\right]\circ F_{\bar\eta}^{-1}(\cdot),
\]
where $\bar\xi_1$, $\bar\xi_2$ and $\bar\eta$ denote the Fréchet means of $\xi_1$, $\xi_2$ and $\eta$, respectively; $(\beta_1,\beta_2)^\top \in \mathbb{R}^2$ are the model parameters; $T_{\bar\xi_j\to\xi_{ji}}$, $j=1,2$ is the optimal transport map from $\bar\xi_j$ to $\xi_{ji}$; and $\{T_{\epsilon_i}\}_{i=1}^N$ are random perturbation maps.

Under the same data-generating setting as in \eqref{generate_multiple}, we compare the RMSEs of the MTDR and GOT models for various values of $\bm{\alpha}^*$, with the results summarized in Table~\ref{rmse_multiple}. For different values of $\bm{\alpha}^*$, the distinct model structures lead to consistently smaller RMSE values for the proposed MTDR model compared with the GOT model in our simulation setting. The case $\bm{\alpha}^* = (1, 0, 0)^\top$ is not considered here, as it corresponds exactly to the $\alpha_1^* = 0$ scenario reported in Table~\ref{rmse_single}. The results of the two methods under the data-generating setting of the GOT model are provided in Supplementary Material S2.

\begin{table}[!htbp]
\centering
\caption{Comparison of RMSE values for different values of $\boldsymbol{\alpha}^*$ under MTDR model setting \eqref{generate_multiple} \((n=m=200)\).
The smaller RMSE value in each row is highlighted in bold.}
    \label{rmse_multiple}
    \begin{tabular}{@{}ccc@{}}
\toprule\noalign{}
\(\bm{\alpha}^*\)   &$RMSE_{MTDR}$&$RMSE_{GOT}$\\
\midrule\noalign{}
      \((0.3,0.35,0.35)^\top\)&\textbf{0.074(0.013)}& 0.080(0.012)\\
      \((0.2,0.2,0.6)^\top\)&\textbf{0.075(0.013)} &0.081(0.012)\\
            \((0,0.5,0.5)^\top\)&\textbf{0.076(0.014)} &0.086(0.012)\\
            \((0,1,0)^\top\)&\textbf{0.076(0.012)}& 0.103(0.009)\\
    \bottomrule
    \end{tabular}
    \end{table}

\section{Data Application}\label{sec:real}
We now turn to a real-data example to showcase the performance and practical advantages of our model. Human mortality has long been a central topic in the fields of demography and aging research. This section uses data from the Human Mortality Database (HMD, \hyperlink{www.mortality.org}{www.mortality.org}), which provides age-at-death distributions for men and women in 34 countries. Our objective is to predict the age-at-death distribution of males in 2010 using two predictors: the age-at-death distributions of females and males in 2005 from the same country. The use of data five years apart is also motivated by practical considerations: For example, in studies of life insurance experience, mortality rates are often reported in five-year groupings \citep{yue2011study}, and researchers have developed models accordingly. Building on this motivation, we attempt to model and predict the target distribution using the following formulation,
\[
\eta_{2010,i}^{m} = T_{\epsilon_i} \# \Gamma(\xi_{2005,i}^{f},\xi_{2005,i}^{m}), \quad \{ \xi_{2005,i}^{f},\xi_{2005,i}^{m}, \eta_{2010,i}^{m} \}_{i=1}^{34}, 
\]
where
\[
\begin{aligned}
\Gamma(\xi_{2005,i}^{f},\xi_{2005,i}^{m})) &= \mathop{\arg\min}_{\omega} \Big\{ \alpha_0 d_{\mathcal{W}}^2(\omega, T_0\#\xi_0) \\
&\quad\quad+ \alpha_1 d_{\mathcal{W}}^2(\omega, T_1\#\xi_{2005,i}^f)+ \alpha_2 d_{\mathcal{W}}^2(\omega, T_2\#\xi_{2005,i}^m) \Big\}. \\
\end{aligned}
\]
We set $\xi_0 = \bar{\eta}_{2010}^m$ (the Fréchet mean of $\{\eta_{2010,i}^m\}$) as the reference distribution. However, other choices of reference distributions were also tested and yielded similar results. 

For comparison, we also consider the GOT model \citep{zhu2023geodesic}, given by
\[
\eta_{2010,i}^{m} = \left[T_{\epsilon_i}\circ(\beta_1\odot T_{\bar\xi_{j_1^*}\to\xi_{j_1^*,i}})\circ(\beta_2\odot T_{\bar\xi_{j_2^*}\to\xi_{j_2^*,i}})\right]\#\bar\eta_{2010}^m,
\]
where $(j_1^*, j_2^*)$ is the true order of the two predictors $\xi_{2005}^m$ and $\xi_{2005}^f$; $\bar\xi_{2005}^m$, $\bar\xi_{2005}^f$ and $\bar\eta_{2010}^f$ denote the Fréchet means of $\xi_{2005,i}^m$, $\xi_{2005,i}^f$ and $\eta_{2005,i}^m$, respectively.

We further compare the proposed MTDR model with the GOT and OT models \citep{ghodrati2022distribution}, where the OT model uses only the 2005 male distribution as a predictor. We implement a leave-one-out cross-validation scheme: for each country $i$, we fit each model using the remaining 33 countries’ samples, and then predict $\eta_{2010,i}^m$. The prediction accuracy is assessed using the average Wasserstein distance (AWD) $:= \frac{1}{34}\sum_{k=1}^{34}d_{\mathcal{W}}(\nu_k,\widehat{\nu}_k)$, with results reported in Table~\ref{loo2}.

\begin{longtable}[]{@{}cccc@{}}
\caption{Leave-one-out cross-validation errors (AWDs) for predicting the 2010 male age-at-death distribution in 34 countries using the MTDR, OT, and GOT models.}\label{loo2}\tabularnewline
\toprule\noalign{}
Model  &  MTDR & OT & GOT\\
\midrule\noalign{}
\endfirsthead
\endhead
\bottomrule\noalign{}
\endlastfoot
AWD  & 0.530  & 0.618 &  0.559 \\ 
\end{longtable}

Table~\ref{loo2} shows that the proposed MTDR model achieves the lowest prediction error among the three methods, outperforming both OT and GOT in this real-data setting. Compared with the OT model, the MTDR model incorporates additional predictors, making its smaller prediction error entirely expected. In fact, even when the MTDR model uses only a single predictor, the 2005 male distribution, it still achieves a substantially lower prediction error. 
This improvement is largely attributable to the non-negligible role played by the reference distribution $\xi_0$ in enhancing predictive accuracy. This finding indicates that incorporating a reference distribution can lead to substantial gains in forecast performance.

In comparison with the GOT model, our method achieves only marginal improvements in prediction error. However, the estimated scalar and map coefficients still provide clear and interpretable insights into the underlying regression structure. Following the explanatory framework proposed by \citet{ghodrati2022distribution}, these map coefficients quantify changes in the shape of the predicted distribution; more formally, they characterize how the probability mass of the predicted distribution is rearranged, while scalar coefficients can be viewed as quantifying the relative contribution of each predictor distribution to the target distribution. 

\begin{figure}
  \centering
  \begin{subfigure}{0.8\textwidth}
    \centering
    \includegraphics[width=\textwidth]{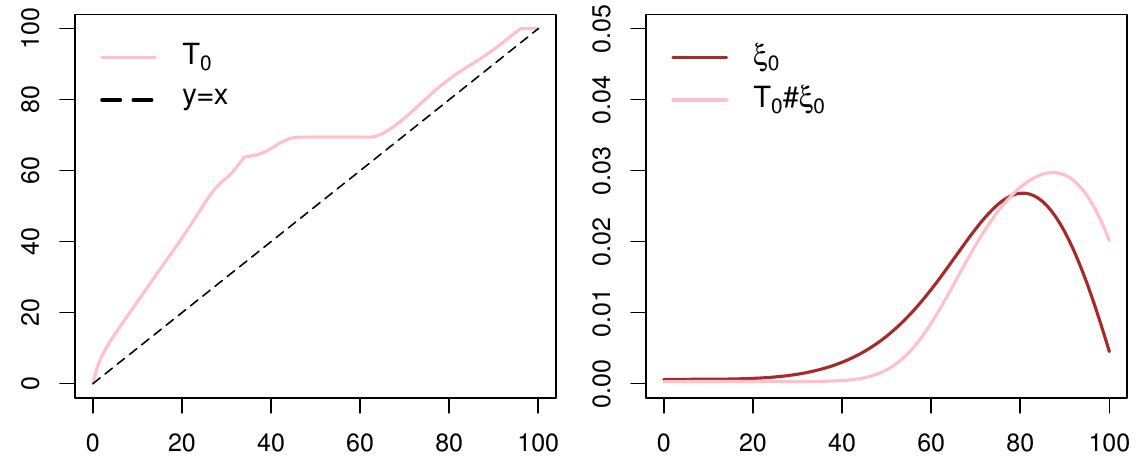}
    \caption{$T_0$}
  \end{subfigure}

  \begin{subfigure}{0.8\textwidth}
    \centering
    \includegraphics[width=\textwidth]{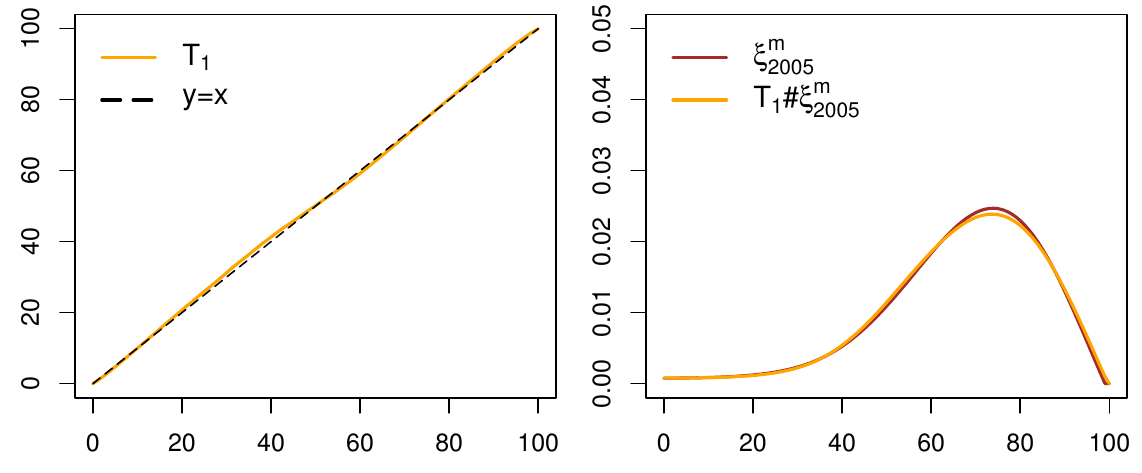}
    \caption{$T_1$}
  \end{subfigure}

  \begin{subfigure}{0.8\textwidth}
    \centering
    \includegraphics[width=\textwidth]{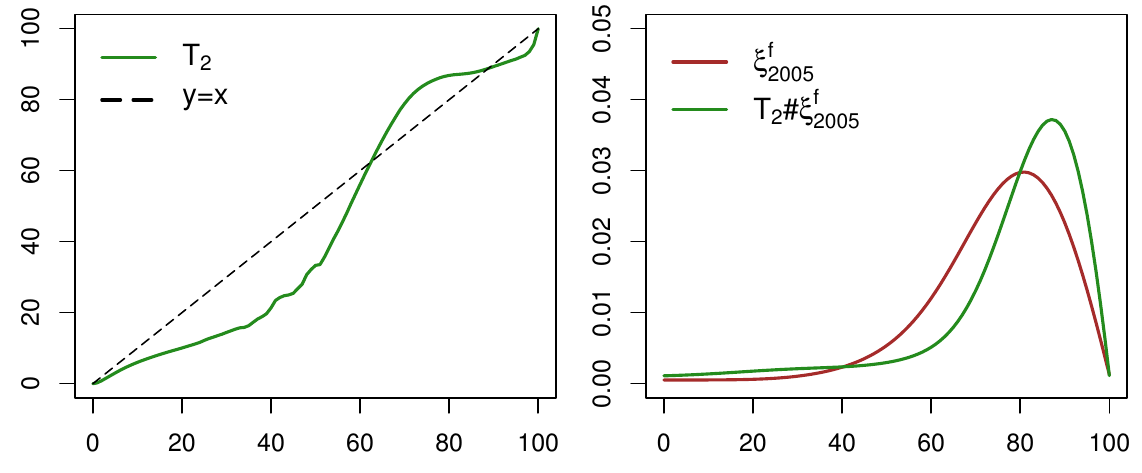}
    \caption{$T_2$}
  \end{subfigure}
   \vspace{-2mm}
   \caption{\label{bul}Estimated transport maps (left column) and their effects on the corresponding predictor distributions (right column) for the case of Bulgaria under the leave-one-out scheme. The solid lines in the left column represent the estimated maps \(T_k, (k=0,1,2)\), and the dashed lines indicate the identity map \(y = x\).}
\end{figure}

As an illustrative example, we consider the case of Bulgaria. Under the leave-one-out scheme, the estimated scalar coefficients are \(\widehat{\alpha}_0 = 0.08\), \(\widehat{\alpha}_1 = 0.83\), and \(\widehat{\alpha}_2 = 0.09\). The corresponding estimated transport maps \(T_0\), \(T_1\) and \(T_2\) are shown in the first column of Figure~\ref{bul}, while the second column shows how these maps modify the shape of the associated predictor distributions. Since \(\xi_0\) is an artificially chosen reference distribution and different choices of \(\xi_0\) would yield different \(T_0\), we focus directly on the composite \(T_0 \# \xi_0\). For \(T_1\), as seen in the left panel of Figure~\ref{bul}(b), the map is very close to the identity line \(y = x\), implying that \(T_1 \# \xi_{2005}^m\) is nearly identical to \(\xi_{2005}^m\). 

\begin{figure}
    \centering
    \includegraphics[width=0.8\textwidth]{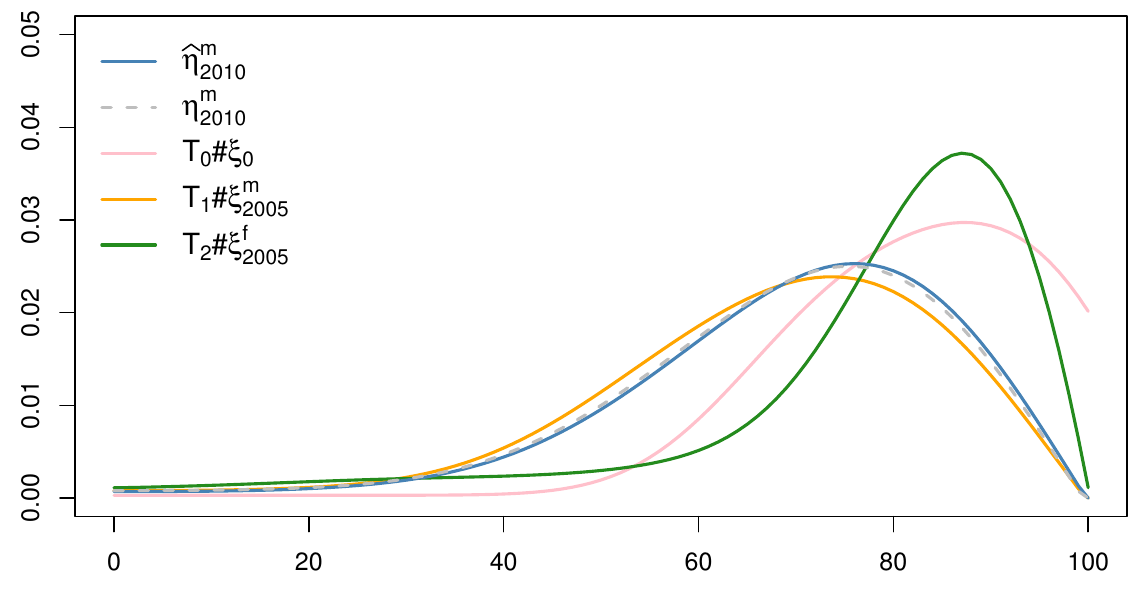}
    \caption{\label{Bul}For Bulgaria, the fitted target distribution $\widehat{\eta}_{2010}^m$ (blue),
the observed target distribution $\eta_{2010}^m$ (gray, dotted),
and the transformed predictor distributions
$T_0\#\xi_0$ (pink), $T_1\#\xi_{2005}^m$ (orange), and $T_2\#\xi_{2005}^f$ (green).
}
\end{figure}

Building on the above interpretation for Bulgaria in Figure~\ref{bul}, we further illustrate the relationship between the transformed predictor distributions and the fitted target distribution. Figure~\ref{Bul} displays the fitted distribution $\widehat{\eta}_{2010}^m$ obtained from the estimated coefficients, the observed target distribution $\eta_{2010}^m$, and the three transformed predictor distributions $T_0\#\xi_0$, $T_1\#\xi_{2005}^m$ and $T_2\#\xi_{2005}^f$. The fitted distribution $\widehat{\eta}_{2010}^m$ corresponds to the weighted Fréchet mean of $T_0\#\xi_0$, $T_1\#\xi_{2005}^m$ and $T_2\#\xi_{2005}^f$, with weights given by the corresponding estimated scalar coefficients \(\widehat{\alpha}_0 = 0.08\), \(\widehat{\alpha}_1 = 0.83\), and \(\widehat{\alpha}_2 = 0.09\). Given the relatively larger weight on $T_1\#\xi_{2005}^m$, the fitted distribution is therefore closer to this transformed predictor.

This example highlights how the proposed model not only achieves accurate prediction, but also provides interpretable insights into the role of each predictor distribution in shaping the target distribution.










\newpage
\phantomsection\label{supplementary-material}
\bigskip

\begin{center}

{\large\bf SUPPLEMENTARY MATERIAL}

\end{center}

\begin{description}
\item[Comparison under GOT model setting for single predictor.]
\item[Comparison under GOT model setting for multiple predictors.]
\item[Proof.]

\end{description}

\bibliography{ref}

@article{yue2011study,
  title={A study of incidence experience for Taiwan life insurance},
  author={Yue, Ching-Syang Jack and Huang, Hong-Chih},
  journal={The Geneva Papers on Risk and Insurance-Issues and Practice},
  volume={36},
  number={4},
  pages={718--733},
  year={2011},
  publisher={Springer}
}

@article{agueh2011barycenters,
  title={Barycenters in the Wasserstein space},
  author={Agueh, Martial and Carlier, Guillaume},
  journal={SIAM Journal on Mathematical Analysis},
  volume={43},
  number={2},
  pages={904--924},
  year={2011},
  publisher={SIAM}
}

@article{chen2024distribution,
  title={Distribution-in-distribution-out Regression},
  author={Chen, Xiaoyu and Fu, Mengfan and Huang, Yujing and Deng, Xinwei},
  journal={arXiv preprint arXiv:2405.11626},
  year={2024}
}

@article{ghosal2025distributional,
  title={Distributional outcome regression via quantile functions and its application to modelling continuously monitored heart rate and physical activity},
  author={Ghosal, Rahul and Ghosh, Sujit K and Schrack, Jennifer A and Zipunnikov, Vadim},
  journal={Journal of the American Statistical Association},
  pages={1--13},
  year={2025},
  publisher={Taylor \& Francis}
}

@article{Bigot2017GPCA,
author = {J{\'e}r{\'e}mie Bigot and Ra{\'u}l Gouet and Thierry Klein and Alfredo L{\'o}pez},
title = {{Geodesic PCA in the Wasserstein space by convex PCA}},
volume = {53},
journal = {Annales de l'Institut Henri Poincaré, Probabilités et Statistiques},
number = {1},
publisher = {Institut Henri Poincaré},
pages = {1--26},
year = {2017},
doi = {10.1214/15-AIHP706},
URL = {https://doi.org/10.1214/15-AIHP706}
}

@article{hron2016simplicial,
  title={Simplicial principal component analysis for density functions in Bayes spaces},
  author={Hron, Karel and Menafoglio, Alessandra and Templ, Matthias and Hr{\r{u}}zov{\'a}, Kl{\'a}ra and Filzmoser, Peter},
  journal={Computational Statistics \& Data Analysis},
  volume={94},
  pages={330--350},
  year={2016},
  publisher={Elsevier}
}

@article{nerini2007classifying,
  title={Classifying densities using functional regression trees: Applications in oceanology},
  author={Nerini, David and Ghattas, Badih},
  journal={Computational Statistics \& Data Analysis},
  volume={51},
  number={10},
  pages={4984--4993},
  year={2007},
  publisher={Elsevier}
}

@article{kokoszka2019forecasting,
  title={Forecasting of density functions with an application to cross-sectional and intraday returns},
  author={Kokoszka, Piotr and Miao, Hong and Petersen, Alexander and Shang, Han Lin},
  journal={International Journal of Forecasting},
  volume={35},
  number={4},
  pages={1304--1317},
  year={2019},
  publisher={Elsevier}
}

@article{Victor2016point,
author = {Victor M. Panaretos and Yoav Zemel},
title = {{Amplitude and phase variation of point processes}},
volume = {44},
journal = {The Annals of Statistics},
number = {2},
publisher = {Institute of Mathematical Statistics},
pages = {771 -- 812},
year = {2016},
doi = {10.1214/15-AOS1387},
URL = {https://doi.org/10.1214/15-AOS1387}
}

@article{Zhang2022,
author = {Zhang, Chao and Kokoszka, Piotr and Petersen, Alexander},
title = {Wasserstein autoregressive models for density time series},
journal = {Journal of Time Series Analysis},
volume = {43},
number = {1},
pages = {30 -- 52},
doi = {https://doi.org/10.1111/jtsa.12590},
url = {https://onlinelibrary.wiley.com/doi/abs/10.1111/jtsa.12590},
eprint = {https://onlinelibrary.wiley.com/doi/pdf/10.1111/jtsa.12590},
year = {2022}
}

@article{Chen2023Wasserstein,
author = {Yaqing Chen and Zhenhua Lin and Müller, Hans-Georg },
title = {Wasserstein Regression},
journal = {Journal of the American Statistical Association},
volume = {118},
number = {542},
pages = {869 -- 882},
year = {2023},
publisher = {Taylor & Francis},
doi = {10.1080/01621459.2021.1956937},
URL = { 
https://doi.org/10.1080/01621459.2021.1956937
},
eprint = { 
https://doi.org/10.1080/01621459.2021.1956937
}
}

@article{chen2019lqd,
  title={LQD-RKHS-based distribution-to-distribution regression methodology for restoring the probability distributions of missing SHM data},
  author={Chen, Zhicheng and Bao, Yuequan and Li, Hui and Spencer Jr, Billie F},
  journal={Mechanical Systems and Signal Processing},
  volume={121},
  pages={655--674},
  year={2019},
  publisher={Elsevier}
}

@article{2016Functional,
    author = {Petersen, Alexander and Müller, Hans-Georg},
    title = {Functional data analysis for density functions by transformation to a Hilbert space},
    volume = {44},
    journal = {The Annals of Statistics},
    number = {1},
    publisher = {Institute of Mathematical Statistics},
    pages = {183 -- 218},
    year = {2016},
    doi = {10.1214/15-AOS1363},
    URL = {https://doi.org/10.1214/15-AOS1363}
}

@article{PETERSEN2022model,
title = {Modeling Probability Density Functions as Data Objects},
journal = {Econometrics and Statistics},
volume = {21},
pages = {159 -- 178},
year = {2022},
issn = {2452-3062},
doi = {https://doi.org/10.1016/j.ecosta.2021.04.004},
url = {https://www.sciencedirect.com/science/article/pii/S245230622100054X},
author = {Petersen, Alexander and Chao Zhang and  Kokoszka, Piotr}
}

@article{ghodrati2022distribution,
  title={Distribution-on-distribution regression via optimal transport maps},
  author={Ghodrati, Laya and Panaretos, Victor M},
  journal={Biometrika},
  volume={109},
  number={4},
  pages={957--974},
  year={2022},
  publisher={Oxford University Press}
}

@article{zhu2023autoregressive,
  title={Autoregressive optimal transport models},
  author={Zhu, Changbo and M{\"u}ller, Hans-Georg},
  journal={Journal of the Royal Statistical Society Series B: Statistical Methodology},
  volume={85},
  number={3},
  pages={1012--1033},
  year={2023},
  publisher={Oxford University Press US}
}

@article{ghodrati2024distributional,
  title={On distributional autoregression and iterated transportation},
  author={Ghodrati, Laya and Panaretos, Victor M},
  journal={Journal of Time Series Analysis},
  volume={45},
  number={5},
  pages={739--770},
  year={2024},
  publisher={Wiley Online Library}
}

@article{zhu2023geodesic,
  author = {Zhu, Changbo and Müller, Hans-Georg},
    title = {Geodesic Optimal Transport Regression},
    journal = {Biometrika},
    pages = {asaf086},
    year = {2025},
    month = {12},
    issn = {1464-3510},
    doi = {10.1093/biomet/asaf086},
    url = {https://doi.org/10.1093/biomet/asaf086},
    eprint = {https://academic.oup.com/biomet/advance-article-pdf/doi/10.1093/biomet/asaf086/65672861/asaf086.pdf},
}

@book{villani2021topics,
  title={Topics in optimal transportation},
  author={Villani, C{\'e}dric},
  volume={58},
  year={2021},
  publisher={American Mathematical Soc.}
}

@article{santambrogio2015optimal,
  title={Optimal transport for applied mathematicians},
  author={Santambrogio, Filippo},
  year={2015},
  publisher={Springer}
}

@article{ambrosio2008gradient,
  title={Gradient flows in metric spaces and in the spaces of probability measures, and applications to fokker-planck equations with respect to log-concave measures},
  author={Ambrosio, Luigi},
  journal={Bollettino dell'Unione Matematica Italiana},
  volume={1},
  number={1},
  pages={223--240},
  year={2008},
  publisher={Unione Matematica Italiana}
}

\end{document}